\documentclass[11pt]{amsart}
\usepackage{amsmath,amsfonts,amssymb}
\usepackage{graphicx}

\title[Quantum Geometry of Measurements]{Quantum Information Geometry in the\\ Space of Measurements}

\author{Warner A. Miller}
\address{Department of Physics, Florida Atlantic University, Boca Raton, FL, 33431}

\begin{document}

\begin{abstract}
We introduce a new approach to evaluating entangled quantum networks using information geometry.  Quantum computing is powerful because of the enhanced correlations from quantum entanglement.  For example, larger entangled networks can enhance quantum key distribution (QKD).   Each network we examine is an $n$-photon quantum state with a degree of entanglement.  We analyze such a state within the space of measured data from repeated experiments made by $n$ observers over a set of identically-prepared quantum states -- a quantum state interrogation in the space of measurements.  Each observer records a $1$ if their detector triggers, otherwise they record a $0$.  This generates a string of $1$'s and $0$'s at each detector, and each observer can define a binary random variable from this sequence.  We use a well-known information geometry-based  measure of distance that applies to these binary strings of measurement outcomes \cite{Rokhlin:1959,Rajski:1960,Zurek:1989}, and we introduce a  generalization of this length to area, volume and higher-dimensional volumes \cite{QUIG:1990}.  These geometric equations are defined using the familiar Shannon expression for joint and mutual entropy \cite{Shannon:1948}.  We apply our approach to three distinct tripartite quantum states:  the $|GHZ\rangle$ state, the  $|W\rangle$ state,  and a separable state $|P\rangle$.  We generalize a well-known information geometry analysis of a bipartite state to a tripartite state.  This approach provides a novel way to characterize quantum states, and it may have favorable scaling with increased  number of photons.
\end{abstract}

\maketitle

% Include a list of keywords after the abstract 
%\keywords{quantum entanglement, quantum information geometry,  entanglement measures, quantum state tomography, elementary quantum phenomenon}

\section{INTRODUCTION}
\label{sec:intro}  % \label{} allows reference to this section

\begin{figure} [htbp]
\centering
   \includegraphics[height=4in]{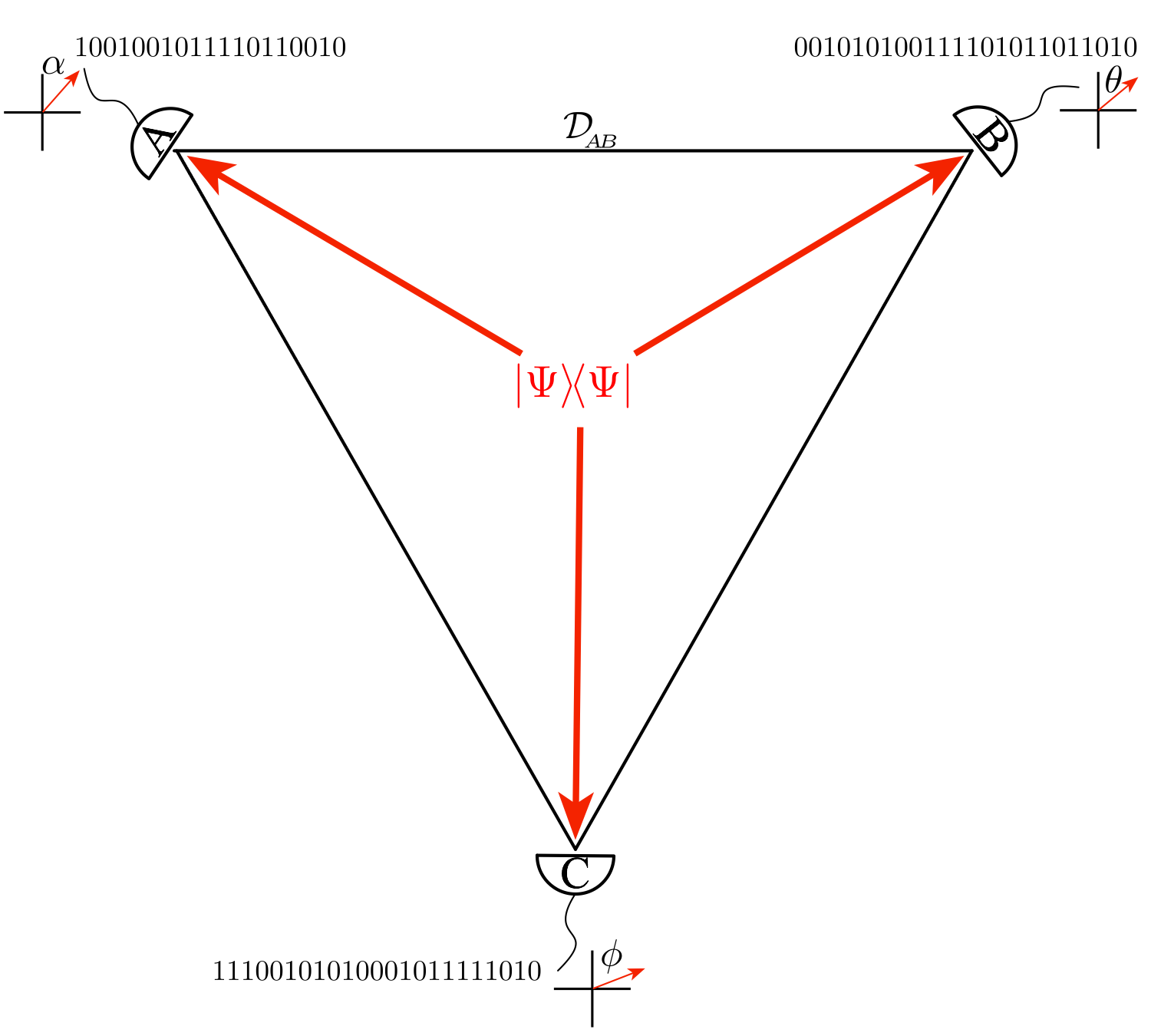}
   \caption{{\em An illustration of a simplicial geometry representation of a quantum tripartite state consisting of three photons.} The geometry of the triangle emerges from a space of measurements that are parameterized by the six parameters of the detectors of our three observers.  The parameters are the two angles on Bloch sphere characterizing each detector.  In this manuscript, and for clarity of presentation only, we restrict our three observers, Alice ($A$), Bob ($B$) and Charlie ($C$), to measurements made within the subspace linearly polarized light (the equator of the Bloch sphere).  This space of measurement can be spanned by three angular parameters, one for each of the three detectors, $\alpha$, $\beta$ and $\gamma$. With the choice of the detectors made, the three observers record a $1$ if their detector gets triggered otherwise they record a $0$.  We analyze the entropy of each of these binary strings as well as calculate the mutual and conditional entropy of pairs and triplets of the detectors. The geometric quantities are functions of these entropies. Therefore, the information geometry of this triangle $\overline{ABC}$ is determined by;  (1) the quantum state $|\Psi\rangle$,  and (2) the three angles, $\alpha$, $\beta$ and $\gamma$ defining orientation of Alice, Bob and Charlie's detectors; respectively. We can calculate the length of each edge,  and the area of the triangle. This will be discussed in Sec.~\ref{sec:G}. 
 \label{fig:QN} 
   }
   \end{figure}

``No elementary quantum phenomenon is a phenomenon until it is brought to close by an irreversible act of amplification.'' This Niels Bohr-inspired quantum adage of John Archibald Wheeler, together with the Principle of Complementarity, is at the very heart of Wheeler's It-from-Bit framework \cite{Wheeler:1990}. In this manuscript, we explore entanglement networks within this information-centric approach. The quantum network we consider in this manuscript is a quantum state with $n$ photons with varying degrees of entanglement.  Observers examine the space of measured data from repeated experiments on a set of identically-prepared quantum states.  Each observer records a $1$ if their detector triggers, otherwise a "0" is recorded.  This generates a string of $1$'s and $0$'s at each detector as illustrated in Fig.~\ref{fig:QN}.  The string of numbers can be represented by a binary random variable.  The observers may have more than one detector, and therefore each observer may acquire more than one binary random variable. Once these random variables are formed, we can apply an information geometry measure of distance, area, volume and n-volumes to the network of observers \cite{Rokhlin:1959,Rajski:1960,QUIG:1990,Zurek:1989}. These measures are defined using the familiar Shannon expression for mutual and conditional entropy \cite{Shannon:1948}.  This will be discussed in Sec.~\ref{sec:G}.  In Sec.~\ref{sec:3},  we apply our approach  to three distinct tripartite quantum states: the $|GHZ\rangle$ state, the $|W\rangle$ state, and a separable state $|P\rangle$.  This novel approach provides us with a natural generalization an information geometry model of Bell's inequality for a bipartite singlet state to a similar analysis tripartite states \cite{Schumacher:1991}.  We provide a brief review of this 2-photon geometry in Sec.~\ref{sec:S} and its generalization in Sec.~\ref{sec:O}. 

The power of quantum computing stems from a network of quantum entanglement, and larger networks can enhance quantum key distribution (QKD)\cite{Jozsa:2003,Piani:2009,Regula:2016}.  Consequently, a primary focus of this field is to find a scalable method to characterize a quantum state, e.g. a measure of its entanglement, and entanglement quality.  Such measures have been elusive \cite{Regula:2016,Horodecki:2009}.  Quantum state tomography is impractical as it involves analyzing an exponentially large matrix. This difficulty is not so surprising as the very power of quantum computing lies with this exponential scaling.  We seek a scalable information geometry entanglement measure that is grounded solely upon the space of measured data from repeated experiments --- a quantum state interrogation in the space of measurements.   This It-from-Bit  approach is based on a projection of the quantum world by measurement, (i.e. an irreversible act of amplification) onto a classical world of bit strings of data (bits). After all, any quantum information processing system must ``meet" the classical world of information to communicate its information content.  It is this network  of detectors that we analyze within the ``It-from-Bit" framework.  The usual conceptual ambiguities that may accompany the quantum measurement process are minimized within this approach; nevertheless, the non-classical and non-intuitive features of the quantum remain.  The uniqueness of quantum phenomenon are now encoded in the correlations of our observers bits.  The two questions we ask are: ``{\em Can information geometry provide us with a better understanding of the entanglement structure and function in quantum networks?}''; and ``{\em Can this information geometry approach scale favorably with larger multipartite systems?}''.

Scalability is the single salient sign signaling a superior quantum strategy. Quantum information processing is driven by exponential scaling, and this must be a concern for any approach to measure or characterize entanglement.  For this reason, our approach will try to emulate the exponential scaling suggested by Vedral \cite{Vedral:1997}.  The scalability of this approach through the Quantum Sanov's Theorem \cite{Vedral:1997} shows that the fidelity of distinguishing two quantum density matrices  $\rho_2$ from $\rho_1$ improves exponentially with the number of measurements, $N$,
\begin{equation}\label{eq:Sanov}
\left(
\begin{array}{c}
Fidelity\ of\ \rho_1 \rightarrow \rho_2\\
with\ N\ measurements
\end{array}
\right)
= 1-e^{-N S\left(\rho_1||\rho_2\right)}.
\end{equation} 
Here, $S(\rho_1||\rho_2) = Tr(\rho_1 \log \rho_1 - \rho_1 \log \rho_2)$ is a relative entropy.  We are animated by this ``information thermodynamic" structure, although we are far from making progress on this front. We outline here an approach to the answers to our two questions that is based on quantum state interrogation in a space of measurements.  We introduce potential measures that utilize quantum information geometry, and look forward in the future to verify if our ``large $N$'' thermodynamic-like collection of bits from measurements share the structure of the exponential scaling suggested by Vedral \cite{Vedral:1997}.   

\section{Outline of Information Geometry: From Distances to Area and Volumes}
\label{sec:G}

Information geometry is defined through the entropy of a network of random variables.  For example if we examine the binary outcomes of Alice's detector in Fig.~\ref{fig:QN}, we can obtain the probability distribution for Alice's ($A$) detector labeled $\alpha$. In particular, we  separately summing the number of times the detector fires and gives a  $1$, and the number non-detections $0$ that she measured. We then divide each of these by the total number of measurements.  In this way we can assign a binary random variable, $A$,  to Alice's 19 measurements (Fig.~\ref{fig:QN}).  
\begin{equation}
A = \left\{  
\begin{array}{lll}
0 & with\ probability & 9/19\\
1 & with\ probability & 10/19
\end{array}
\right..
\end{equation}
Whereas probability measures uncertainty about the occurrence  of a single event, entropy provides a measure the uncertainty of a collection of events. If $X_i$ is a $s$-state random variable, then 
\begin{equation}
\label{eq:entropy}
\left( \begin{array}{c} Entropy\\ of\ X_i\end{array} \right) = H_{X_i} := - \sum_{\chi_i=1}^s  p(x_i) \log p(x_i).
\end{equation}
Here $p(x_i)=p(X\!=\!x_i)$ is the probability that the random variable has the value $x_i$.  In this manuscript, we use only binary random variables ($s=2$).  The entropy is the largest when our uncertainty of the value of the random variable is complete (e.g. uniform distribution of probabilities), and the entropy is zero if the random variable always takes on the same value, 
\begin{equation}
0 \le H_{X_i} \le \log{(s)}.
\end{equation}
In this sense, entropy is a measure of our ignorance.  We will make use of the mutual entropy and conditional entropy over an ensemble of random variables. The mutual entropy is defined over the joint probability distributions,
\begin{equation}
H_{ABC} = -\sum_{i,j,k} p(a_i,b_j,c_k) \log p(a_i,b_j,c_k), 
\end{equation}
and the conditional entropy is defined over conditional probability distributions, 
\begin{align}
\label{eq:centropy}
H_{A|B} &= -\sum_{i,j} p(b_j) \log{p(a_i|b_j)},\\
H_{A | B C} &=  -\sum_{i,j,k} p(b_j,c_k) \log p(a_i|b_j,c_k).
\end{align}
Here the probability that $A=a_i$, $B=b_j$ and $C=c_k$ is the joint probability $p(a_i,b_j,c_k)$, and the probability that $A=a_i$ given that we know a priori that  $B=b_j$ and $C=c_k$ is the conditional probability $p(a_i|b_j,c_k)$, and these are related by
\begin{equation}
\label{eq:probobility}
p(a_i|b_jc_k) = \frac{p(a_i,b_j,c_k)}{p(b_j,c_k)}.
\end{equation}
We use use an extension of the Shannon-based information distance defined by  Rokhlin\cite{Rokhlin:1959} and Rajski\cite{Rajski:1960}, 
\begin{equation}\label{eq:length}
\left(
\begin{array}{c}
Length\ of\\
Edge\ AB
\end{array}
\right) =
{\mathcal D}_{AB} := H_{A|B}+H_{B|A} = 2 H_{AB} -H_A-H_B,
\end{equation}
to construct a geometric triangle from these measurement outcomes. Here in Eq.~\ref{eq:length}, 
$A$ and $B$ are binary random variables derived from a joint probability distribution $p(A\!=\!a_i,B\!=\!b_i)=p(a_i,b_i)$ with $i\in\{i,s\}$.  This information distance has desirable properties. 
\begin{enumerate}
\item It is constructed to be symmetric, ${\mathcal D}_{AB}={\mathcal D}_{AB}$.
\item It obeys the triangle inequality, ${\mathcal D}_{AB} \le {\mathcal D}_{Ac}+{\mathcal D}_{CB}$. 
\item It is nonnegative, ${\mathcal D}_{AB}\le 0$, and equal to $0$ when  $A$``=''$B$.
\end{enumerate}
Furthermore, if $A$ and $B$ are uncorrelated to each other then, 
\begin{equation}
\label{eq:id} 
{\mathcal D}_{AB} = 2\left(H_A+H_B\right) - H_A-H_B=H_A+H_B, 
\end{equation}
and ${\mathcal D}_{AB}$ is bounded,
\begin{equation}
0 \le {\mathcal D}_{AB} \le H_A+H_B \le 2 \log s.
\end{equation}

In addition to the three edge lengths of the triangle in Fig.~\ref{fig:QN},  we can assign an information area that we developed earlier with Caves, Kheyfets, Lloyd, Miller, Schumacher and Wotters \cite{QUIG:1990},
\begin{eqnarray}
\label{eq:area}
\begin{array}{rl}
{\mathcal A}_{ABC}   &:=   H_{A|BC}H_{B|CA}+H_{B|CA}H_{C|AB}+H_{C|AB}H_{A|BC}\\
& =  3 H_{ABC}^2 - 2 (H_{AB}+H_{BC}+H_{AC}) H_{ABC} + (H_{AC}H_{BC}+H_{AB}H_{AC}+H_{AB}H_{BC}).
\end{array}
\end{eqnarray}
This can be generalized to higher-dimensional simplexes, e.g.  the information volume for a tetrahedron can be defined as, 
\begin{equation}
\label{eq:volume}
\begin{array}{ll}
{\mathcal V}_{ABCD}& := H_{A|BCD}H_{B|CDA}H_{C|DAB} +H_{B|CDA}H_{C|DAB}H_{D|ABC}\\
&\ \ +H_{C|DAB}H_{D|ABC}H_{B|CDA}+H_{D|ABC}H_{A|BCD}H_{B|CDA}.
\end{array}
\end{equation}
For classical probability distributions these formulas are well defined and have all of the requisite symmetries, positivity, bounds and structure usually required for such formulae.   In particular, 
\begin{equation}
0 \leq  {\mathcal D}_{AB} \leq H_A + H_B \leq \underbrace{2 \log{s}}_{\hbox{$s$-state r.v's}},
\end{equation}
and
\begin{equation}
\label{eq:areabounds}
0 \leq {\mathcal A}_{ABC}  \leq 3 \left( \log{s}\right)^2,
\end{equation}
with their minimum values taken when the random variables are completely correlated,  and their maximum values obtained when the random variables are completely uncorrelated. We have shown that for classical probability distributions these formulas are well defined and have all requisite symmetries, positivity, bounds and structure required for such formulae \cite{QUIG:1990}.
However, for probability distributions based on measurements of quantum systems these assumptions must be weakened, and triangle inequalities may be violated.  We will outline such an example in Sec.~\ref{sec:S}. 

We are confident that these novel formulae, Eqns.~\ref{eq:area}-\ref{eq:volume}, can provide a new characterization of quantum states and their degree of entanglement. We may need far fewer measurements than one would think to distinguish these states.  In particular, the Quantum Sanov's Theorem \cite{Vedral:1997} shows that the fidelity of distinguishing two quantum states $\rho_2$ from $\rho_1$ improves exponentially with the number of measurements, $N$ as seen in Eq.~\ref{eq:Sanov}.  This ``information thermodynamic'' feature may lay credence our scalability assumption; however this requires further investigation. 

\section{Quantum State Interrogation of the Singlet State: A Review }
\label{sec:S}

\begin{figure} [ht]
\centering
   \includegraphics[height=2.5in]{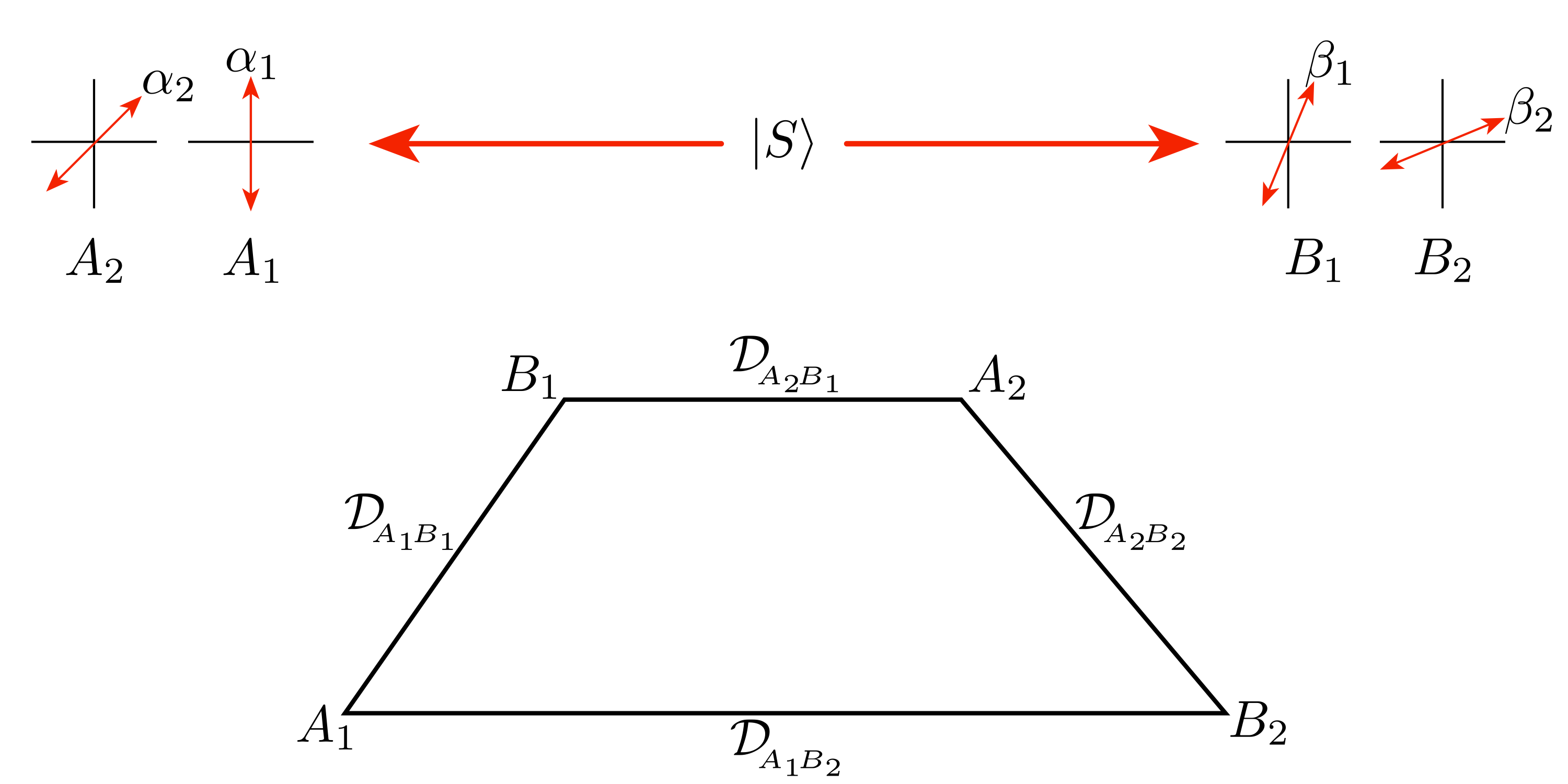}
   \caption{We illustrate here the information geometry of a singlet state analyzed by Schumacher \cite{Schumacher:1991}. There are two observers, Alice and Bob, that are detecting the 2-photons from the singlet state $|S\rangle$. Alice has two detectors, one linear polarizer rotated an angle $\alpha_1$ away from vertical,  the other detector is rotated an angle $\alpha_2$, and similarly for Bob.  An ensemble of singlet states are prepared and Alice and Bob randomly choose one or the other detector. This leads after many measurements to four binary random variables, $A_1$, $A_2$, $B_1$ and $B_2$. in the bottom of the figure, we show the quadrilateral formed by these for random variables. We can use Eq.~\ref{eq:length} to calculate the four distances ${\mathcal D}$'s shown on the edges. We cannot connect the diagonals as they are mutually exclusive; therefore, we can not define an information area. 
 \label{fig:S} 
   }
   \end{figure}

For a bipartite quantum system we can explore the information geometry through the distance formula given in Eq.~\ref{eq:length}. One can look for a relationship between the entanglement and its geometry. Based on this approach Schumacher examined the relationship between the violation of the Bell inequality for a singlet state and the triangle inequality in information geometry \cite{Schumacher:1991}. This is illustrated in Fig.~\ref{fig:S}.  Here we review his results in detail as this is the simplest non-trivial application of this formalism.  We provide many identical copies of a singlet state,
\begin{equation}
\label{eq:S}
|S\rangle = \frac{1}{\sqrt{2}} \left( |\updownarrow \updownarrow\rangle + |\leftrightarrow \leftrightarrow \rangle \right),
\end{equation}
and two observers Alice and Bob as shown in Fig.~\ref{fig:S}.  Alice receives the photon propagating to the left, and Bob receives the photon traveling to the right. Alice choses randomly one of two detectors. Alice's first detector, $\langle \alpha_1 |$,  is a linear polarizer rotated clockwise from the vertical state $|\updownarrow\rangle$ by an angle  $\alpha_1$, and her second detector is rotated by an angle $\alpha_2$.  Similarly, Bob's first and second detectors are rotated by $\beta_1$ and $\beta_2$; respectively. 
We perform this calculation for arbitrary angles and for a symmetric photon singlet state.  Schumacher considered an anti-symmetric spin $1/2$ singlet state and used the angles  $\alpha_1=0$, $\alpha_2=\pi/4$, $\beta_1=\pi/8$ and $\beta_2=3\pi/8$. 
When we project the state in Eq.~\ref{eq:S} on these four detectors,  we find the eight probabilities for the measurement outcomes of $A_1$, $A_2$, $B_1$ and $B_2$, and they  are all equally likely.  
\begin{equation} 
\label{eq:p1}
p(A_1=0)  = p(A_1=1) = 
p(A_2=0) =  p(A_2=1) =
p(B_1=0) =  p(B_1=1) = 
p(B_2=0) = p(B_2=1) = \frac{1}{2}.
\end{equation} 
We then calculate two consecutive local measurements on each pair of detectors in order to determine the four sets of conditional probabilities ($A_1$-$B_1$, $A_1$-$B_2$, $A_2$-$B_1$ and $A_2$-$B_2$) In particular for the two detectors $A_1$ and $B_1$, 
\begin{equation} \label{eq:pAB}
\begin{array}{lr}
p(A_1=0|B_1=0) = \cos^2{(\beta_1-\alpha_1)}, & p(A_1=1|B_1=0) =  \sin^2{(\beta_1-\alpha_1)},\\
p(A_1=0|B_1=1) = \sin^2{(\beta_1-\alpha_1)},  & p(A_1=1|B_1=1) = \cos^2{(\beta_1-\alpha_1)}, \\
p(B_1=0|A_1=0) = \cos^2{(\beta_1-\alpha_1)}, & p(B_1=1|A_1=0) = \sin^2{(\beta_1-\alpha_1)}, \\
p(B_1=0|A_1=1) = \sin^2{(\beta_1-\alpha_1)}, & p(B_1=1|A_1=1) = \cos^2{(\beta_1-\alpha_1)}.
\end{array} 
\end{equation}
The conditional probability expressions for $A_1$ and $B_2$ are the same as those in Eq.~\ref{eq:pAB} except we must substitute $\beta_1 \rightarrow\beta_2$. For $A_2$ and $B_1$ we modify Eq.~\ref{eq:pAB} with  $\alpha_1 \rightarrow\alpha2$, and for  $A_2$ and $B_2$ we substitute both angles, i.e. $\alpha_1 \rightarrow\alpha2$ and $\beta_1 \rightarrow\beta_2$.
The joint probabilities can be recovered from Eqs.~\ref{eq:p1}-\ref{eq:pAB} using  
Eq.~\ref{eq:joint}.  In this example, each joint probability is just half its conditional probability. 

We are now in a position to use Eq.~\ref{eq:entropy} and Eq.~\ref{eq:centropy} to calculate the entropy, the conditional entropy as well as the information distance in Eq.~\ref{eq:length}. We find that the entropies are maximal and consistent with the complete uncertainty in the outcome of each measurement by $A$ or $B$.  This is  reflected in Eq.~\ref{eq:p1}) where
\begin{equation}
\label{eq:HAHB}
H_A = H_B= -\frac{1}{2} \log \frac{1}{2}- \frac{1}{2} \log \frac{1}{2} = 1.
\end{equation}  
The joint entropies are more interesting, and can be obtained from Eq.~\ref{eq:pAB} and Eq.~\ref{eq:joint},
\begin{align}
\label{eq:je2}
H_{A_1B_1} & =  1-\sin^2{(\beta_1-\alpha_1)} \log\left(\sin^2{(\beta_1-\alpha_1)}\right) - \cos^2{(\beta_1-\alpha_1)} \log\left(\cos^2{(\beta_1-\alpha_1)}\right),\\
H_{A_1B_2} &=   1- \sin^2{(\beta_2-\alpha_1)} \log\left(\sin^2{(\beta_2-\alpha_1)}\right) -  \cos^2{(\beta_2-\alpha_1)} \log\left(\cos^2{(\beta_2-\alpha_1)}\right),\\
H_{A_2B_1} &=   1- \sin^2{(\beta_1-\alpha_2)} \log\left(\sin^2{(\beta_1-\alpha_2)}\right) -  \cos^2{(\beta_1-\alpha_2)} \log\left(\cos^2{(\beta_1-\alpha_2)}\right),\\
H_{A_2B_2} &=   1- \sin^2{(\beta_2-\alpha_2)} \log\left(\sin^2{(\beta_2-\alpha_2)}\right) -  \cos^2{(\beta_2-\alpha_2)} \log\left(\cos^2{(\beta_2-\alpha_2)}\right). 
\end{align}
We find the four lengths of the quadrilateral in the lower part of Fig.~\ref{fig:S} using  Eqs.~\ref{eq:id}, \ref{eq:je2} and \ref{eq:HAHB},
\begin{eqnarray*}
\label{eq:4dist}
{\mathcal D}_{A_1B_1} & = &  -2\sin^2{(\beta_1-\alpha_1)} \log\left(\sin^2{(\beta_1-\alpha_1)}\right) -2 \cos^2{(\beta_1-\alpha_1)} \log\left(\cos^2{(\beta_1-\alpha_1)}\right),\\
{\mathcal D}_{A_1B_2} & = & -2\sin^2{(\beta_2-\alpha_1)} \log\left(\sin^2{(\beta_2-\alpha_1)}\right) -2 \cos^2{(\beta_2-\alpha_1)} \log\left(\cos^2{(\beta_2-\alpha_1)}\right),\\
{\mathcal D}_{A_2B_1} & = & -2\sin^2{(\beta_1-\alpha_2)} \log\left(\sin^2{(\beta_1-\alpha_2)}\right) - 2\cos^2{(\beta_1-\alpha_2)} \log\left(\cos^2{(\beta_1-\alpha_2)}\right),\\
{\mathcal D}_{A_2B_2} & = & -2\sin^2{(\beta_2-\alpha_2)} \log\left(\sin^2{(\beta_2-\alpha_2)}\right) - 2\cos^2{(\beta_2-\alpha_2)} \log\left(\cos^2{(\beta_2-\alpha_2)}\right). 
\end{eqnarray*}

 If the quadrilateral formed by the four detectors as illustrated in Fig.~\ref{fig:S} was embedded in a Euclidean surface, then the direct route $A_1 \rightarrow B_2$ should always be greater than or equal to the indirect route $A_1 \rightarrow B_1 \rightarrow A_2 \rightarrow B_2$,   
 \begin{equation}
 {\mathcal D}_{A_1B_2} \le  {\mathcal D}_{A_1B_1} +  {\mathcal D}_{A_2B_1} +  {\mathcal D}_{A_2B_2}.
 \end{equation}
 However, Schumacher showed that this triangle inequality is violated for certain angles \cite{Schumacher:1991}. For our symmetric 2-photon singlet state we obtain the same violation. In particular, we find a maximal violation within a symmetric sub-space where three of the pairwise detectors have the same difference in their relative angular settings, whilst the relative angular setting  between the direct connection between $A_1$ and $B_2$ is three times larger, 
 \begin{equation}
\beta_1-\alpha_1 = \beta_1=\alpha_2 = \beta_2-\alpha_2 \approx  0.15234,
\end{equation} 
and therefore,
\begin{equation}
\beta_2-\alpha_1 = 3 (\beta_1-\alpha_1). 
\end{equation}
This yields a violation in the triangle inequality which Schumacher suggests is an information geometry realization of the violation of the Bell Inequality for the maximally entangled singlet state $|S\rangle$. Here, 
\begin{equation}
\underbrace{{\mathcal D}_{A_1B_2}}_{1.42252}    \not\le  \underbrace{{\mathcal D}_{A_1B_1} +  {\mathcal D}_{A_2B_1} +  {\mathcal D}_{A_2B_2}}_{0.948753}.
\end{equation}
While we can further explore this bipartite example, it will be reported in our future work. Nevertheless, this bipartite example of information geometry motivates us to begin to explore tripartite states.  We will outline an analysis of two entangled tripartite states and a seperable state in the next section.

\section{Extending from Bipartite to Tripartite Quantum Networks}
\label{sec:3}

In this section we examine the information geometry of  a tripartite state function --- a generalization of the bipartite work of Schumacher that was discussed in the previous section \cite{Schumacher:1991}.  We will focus on the information geometry for  three distinct states, one seperable quantum state and two well studied entangled states. In particular, we examine the following three states:
\begin{enumerate} 
\item $|\psi_3\rangle = |GHZ\rangle = \frac{1}{\sqrt{2}} \left( |\updownarrow\updownarrow\updownarrow\rangle + | \leftrightarrow\leftrightarrow\leftrightarrow \rangle \right)$;
\item $|\psi_2\rangle = |W\rangle = \frac{1}{\sqrt{3}} \left(  |\updownarrow \updownarrow \leftrightarrow\rangle + |\updownarrow \leftrightarrow \updownarrow \rangle + |\leftrightarrow\updownarrow \updownarrow \rangle \right) $; and 
\item $|\psi_1\rangle = |P\rangle = |\updownarrow \updownarrow \updownarrow\rangle$.
\end{enumerate}
In the next three subsections, Sec.~\ref{sec:GHZ}-\ref{sec:P}, we examine the geometry of Fig.~\ref{fig:QN} for each of these states. Once again, we will restrict ourselves to only linear polarization measurements on the equator of the Bloch sphere.  In  Sec.~\ref{sec:O}, we describe a octagonal network for our tripartite system that is the analogue of the quadrilateral in Fig~\ref{fig:S} for the bipartite system of Schumacher \cite{Schumacher:1991}. 

%*************************
\subsection{Quantum State Interrogation: the $|GHZ\rangle$ State}
\label{sec:GHZ} 

We analyze the information geometry of the triangle shown in Fig:~\ref{fig:QN} for the Greenberger, Horne \& Zeilinger (GHZ) tripartite state,
\begin{equation}
\label{eq:GHZ}
|\Psi\rangle \Longrightarrow |GHZ\rangle = \frac{1}{\sqrt{2}} \left( 
	|\updownarrow\updownarrow\updownarrow\rangle +  
	|\leftrightarrow\leftrightarrow\leftrightarrow\rangle \right)
\end{equation}
We will calculate the three edge lengths using the techniques introduced in Sec.~\ref{sec:G} and applied in Sec:~\ref{sec:S}.  We will also calculate the information area Eq.~\ref{eq:area} for this triangle.  This is something we could not do with the bipartite system in Sec.~\ref{sec:S}.  

\begin{figure}[htbp] %  figure placement: here, top, bottom, or page
   \centering
   \includegraphics[width=4in]{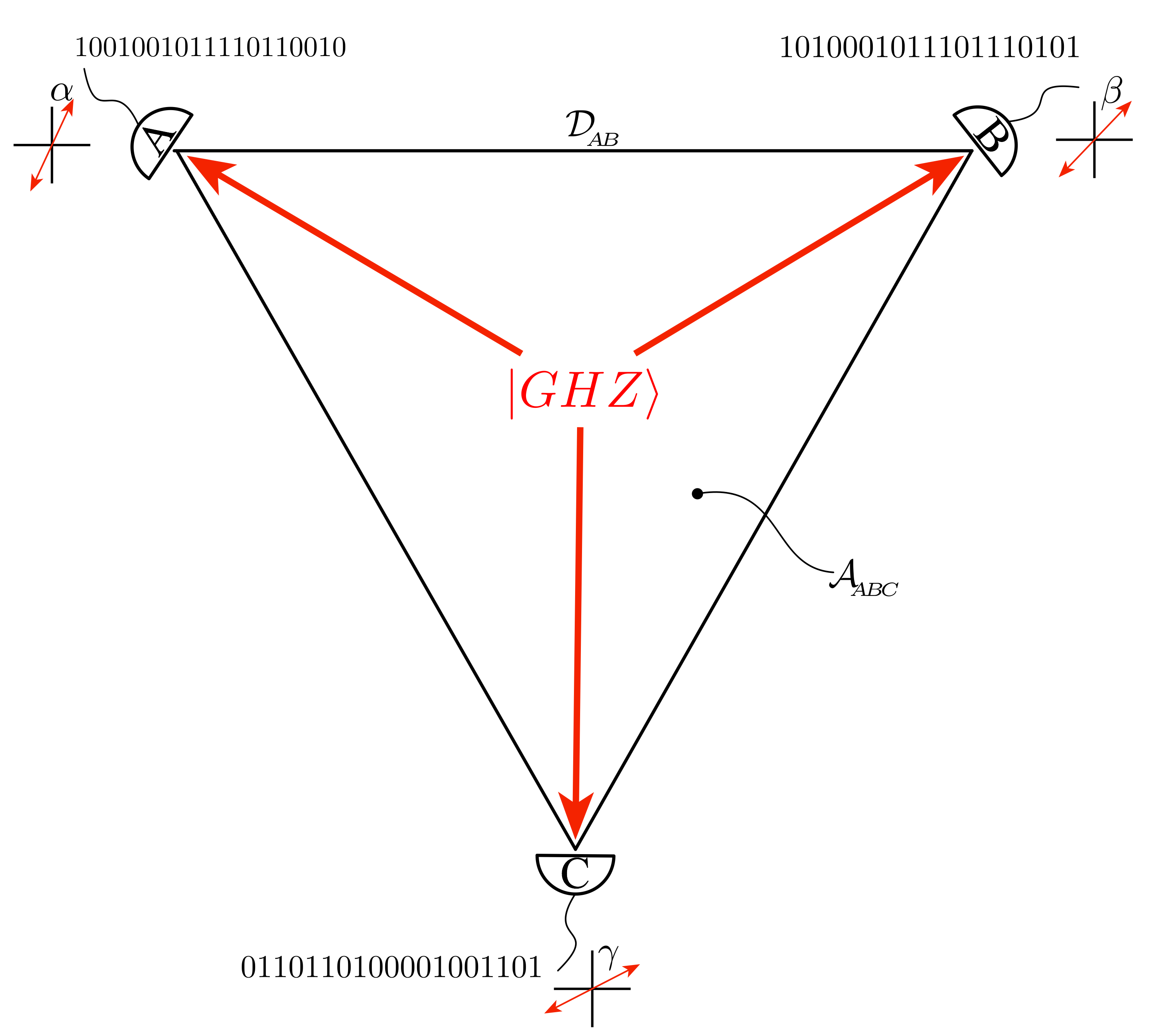} 
   \caption{  {\it The information geometry triangle formed by our three observers, $A$, $B$, and $C$.  These observers share the photons from a $|GHZ\rangle$ state shown in red. Each of the three black edges can be assigned an information length (we show ${\mathcal D}_{AB}$), and the resulting triangle can be assigned an area, ${\mathcal A}_{ABC}^{|GHZ\rangle}$.}
   }
   \label{fig:GHZ}
\end{figure}

 We consider three observers Alice ($A$), Bob ($B$) and Charlie ($C$) as shown in Fig.~\ref{fig:GHZ}. $A$, $B$ and $C$ measure the GHZ state using their choice of detectors, 
 \begin{eqnarray}
 \label{eq:mo}
 M_A  &= &\left(\cos{(\alpha)} \sigma_z +\sin{(\alpha)} \sigma_x\right)\otimes I\otimes I \\
 M_B &= &I\otimes \left(\cos{(\beta)} \sigma_z +\sin{(\beta)} \sigma_x\right) \otimes I\\
 M_C &= &I \otimes I\otimes \left(\cos{(\gamma)} \sigma_z +\sin{(\gamma)} \sigma_x\right);
 \end{eqnarray}
respectively.  Here the $\sigma$'s are the usual Pauli matrices. The probability of $A$ measuring  a photon is
\begin{equation}
\label{eq:prob}
p(A)=tr\left( M_A^\dag M_A  \rho_{\!{}_{GHZ}}\right), 
\end{equation}
where $\rho_{\!{}_{GHZ}} = |GHZ\rangle\langle GHZ|$ is the density matrix for the GHZ state. 
If the initial state was $|GHZ\rangle$ then after $A$'s measurement the state would be left in 
\begin{equation} \label{eq:state}
|GHZ^{A}\rangle = \frac{M_A |GHZ\rangle}{\sqrt{ \langle GHZ| M^\dag_A M_A |GHZ\rangle }}.
\end{equation}
For the remainder of this section we will set $\alpha=0$.

The eight joint probabilities from the three measurements on this entangled state $M_C M_B M_A |GHZ\rangle$ are:
\begin{align}
\label{eq:jointGHZ}
\begin{array}{ll}
p(A=1,B=1,C=1) = \frac{1}{2} \cos^2(\beta)\cos^2(\gamma), &  p(A=0,B=1,C=1) = \frac{1}{2} \sin^2(\beta)\sin^2(\gamma), \\
p(A=1,B=1,C=0) =  \frac{1}{2}\cos^2(\beta)\sin^2(\gamma), & p(A=0,B=1,C=0) =  \frac{1}{2}\sin^2(\beta)\cos^2(\gamma), \\
p(A=1,B=0,C=1) =  \frac{1}{2}\sin^2(\beta)\cos^2(\gamma), & p(A=0,B=0,C=1) =  \frac{1}{2}\cos^2(\beta)\sin^2(\gamma), \\
p(A=1,B=0,C=0) =  \frac{1}{2}\sin^2(\beta)\sin^2(\gamma), & p(A=0,B=0,C=0) =  \frac{1}{2}\cos^2(\beta)\cos^2(\gamma)\\
\end{array}
\end{align}
Tracing these joint probability over each observer yields the pairwise joint probabilities, 
\begin{align}
\begin{array}{ll}
p(A=1,B=1) &= \frac{1}{2}\cos^2(\beta),\\
p(A=1,B=0) &= \frac{1}{2}\sin^2(\beta), \\
p(A=1,C=1) &= \frac{1}{2}\cos^2(\gamma),\\
p(A=1,C=0) &= \frac{1}{2}\sin^2(\gamma),\\
p(A=0,B=1) &= \frac{1}{2}\sin^2(\beta),\\
p(A=0,B=0) &= \frac{1}{2}\cos^2(\beta), \\
p(A=0,C=1) &= \frac{1}{2}\sin^2(\gamma),\\
p(A=0,C=0) &= \frac{1}{2}\cos^2(\gamma),\\
p(B=1,C=1) &= \frac{1}{2}\left( \cos^2(\beta)\cos^2(\gamma) + \sin^2(\beta)\sin^2(\gamma) \right), \\
p(B=1,C=0) &= \frac{1}{2}\left( \cos^2(\beta)\sin^2(\gamma) + \sin^2(\beta)\cos^2(\gamma) \right), \\
p(B=0,C=1) &=  \frac{1}{2}\left( \sin^2(\beta)\cos^2(\gamma) + \cos^2(\beta)\sin^2(\gamma) \right), \\
p(B=0,C=0) &=  \frac{1}{2}\left( \cos^2(\beta)\cos^2(\gamma) + \sin^2(\beta)\sin^2(\gamma) \right). 
\end{array}
\end{align}
Finally, tracing the joint probability over all pairs of observers gives us the six probabilities for the measurement outcomes of $A$, $B$ and $C$ to be 
\begin{equation} \label{eq:p1GHZ}
\begin{array}{lr}
p(A=0) = 1/2, & p(A=1) = 1/2 \\
p(B=0) = 1/2, & p(B=1) = 1/2 \\
p(C=0) = 1/2, & p(C=1) = 1/2
\end{array}.
\end{equation} 
The pairwise conditional probabilities can be recovered from these pairwise joint probabilities since  
\begin{equation} \label{eq:joint}
  p(A=i|B=j) = \frac{p(A=i,B=j)}{p(B=j)}.
\end{equation}
However, since $p(A=i)=p(B=i)=P(C=i)=1/2$ $\forall i\in\{0,1\}$ then the joint probabilities for the $|GHZ\rangle$ state are just half the conditional probabilities. 

We are now in a position to use Eqs.~\ref{eq:entropy}-\ref{eq:centropy} to calculate the entropy, the conditional entropy as well as the information distance in Eq.~\ref{eq:length}.  The entropy of our observers are maximal,
\begin{align}
H_A &= 1,\\
H_B &= 1,\\
H_C &= 1,
\end{align}
and the joint entropy between pairs of our observers are,
\begin{equation}
\label{eq:pe1GHZ}
\begin{array}{rl}
H_{AB} = &  1-\sin^2(\beta) \log(\sin^2(\beta)) -\cos^2(\beta) \log(\cos^2(\beta)),\\
H_{AC} = & 1-\sin^2(\gamma) \log(\sin^2(\gamma)) -\cos^2(\gamma) \log(\cos^2(\gamma)),\\
H_{BC} = & 1-\left( \cos^2(\beta)\cos^2(\gamma) + \sin^2(\beta)\sin^2(\gamma) \right) \log\left( \cos^2(\beta)\cos^2(\gamma) + \sin^2(\beta)\sin^2(\gamma) \right)\\
& -\left( \sin^2(\beta)\cos^2(\gamma) + \cos^2(\beta)\sin^2(\gamma) \right) \log(\left( \sin^2(\beta)\cos^2(\gamma) + \cos^2(\beta)\sin^2(\gamma) \right)).
\end{array}
\end{equation}
Finally, we use Eq.~\ref{eq:joint} to find the joint entropy $H_{ABC}$ of $A$, $B$ and $C$,
\begin{equation}
\label{eq:jeGHZ}
\begin{array}{rl}
H_{ABC} =& 1-\cos^2(\beta)\cos^2(\gamma) \log( \cos^2(\beta)\cos^2(\gamma)) - \cos^2(\beta)\sin^2(\gamma) \log(\cos^2(\beta)\sin^2(\gamma)) \\
                 &{\  \  }  -\sin^2(\beta)\cos^2(\gamma) \log(\sin^2(\beta)\cos^2(\gamma)) - \sin^2(\beta)\sin^2(\gamma) \log(\sin^2(\beta)\sin^2(\gamma)).
                \end{array}
\end{equation}

The three lengths of the edges of the triangle for this GHZ state are derived from Eq.~\ref{eq:length}, and are
\begin{eqnarray}
\label{eq:3l}
{\mathcal D}_{AB} & = &  -2\sin^2{(\beta)} \log\left(\sin^2{(\beta)}\right) -2 \cos^2{(\beta)} \log\left(\cos^2{(\beta)}\right),\\
{\mathcal D}_{AC} & = & -2\sin^2{(\gamma)} \log\left(\sin^2{(\gamma)}\right) -2 \cos^2{(\gamma)} \log\left(\cos^2{(\gamma)}\right),\\\
{\mathcal D}_{BC} & = & -\left( \cos^2(\beta)\cos^2(\gamma) + \sin^2(\beta)\sin^2(\gamma) \right) \log\left( \cos^2(\beta)\cos^2(\gamma) + \sin^2(\beta)\sin^2(\gamma) \right)\\
&& -\left( \sin^2(\beta)\cos^2(\gamma) + \cos^2(\beta)\sin^2(\gamma) \right) \log(\left( \sin^2(\beta)\cos^2(\gamma) + \cos^2(\beta)\sin^2(\gamma) \right)).
\end{eqnarray}

\begin{figure} [ht]
   \includegraphics[height=2 in]{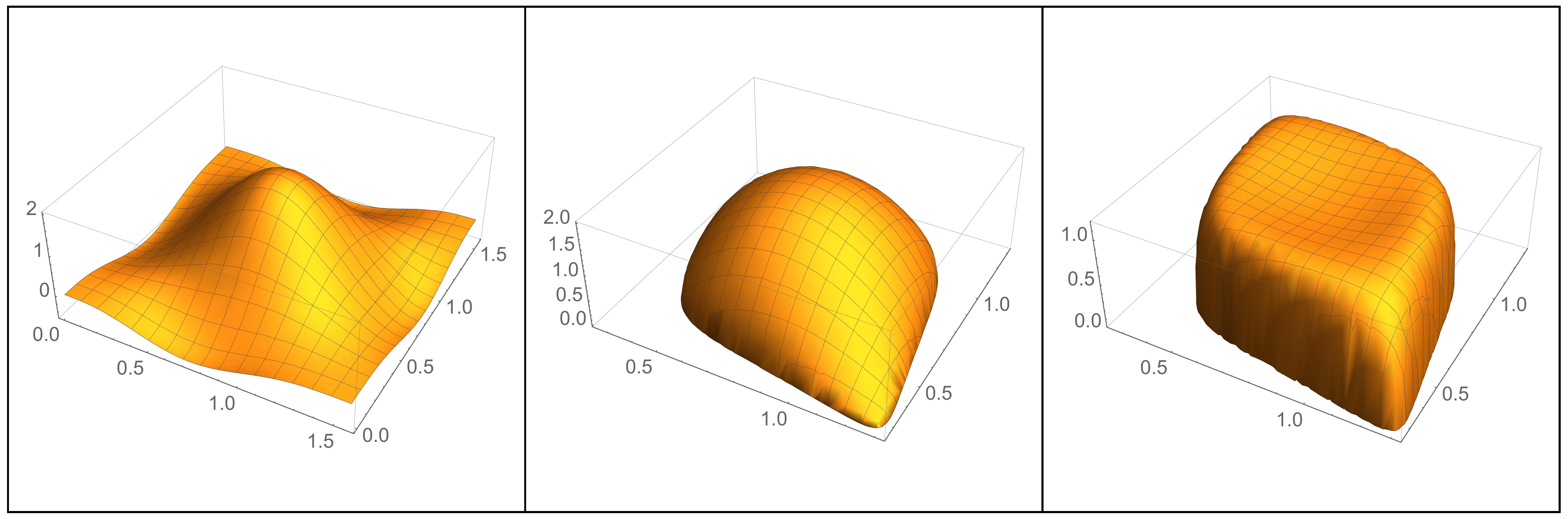}
   \caption
   { \label{fig:GHZ2} 
   The left plot is the information area, ${\mathcal A}_{ABC}^{|GHZ\rangle}(\beta,\gamma)$, the center plot is the Euclidean area (${\mathcal A}_E$) based on the three information distances in Eq.~\ref{eq:3l}, and the right plot is the ratio of these two areas. This ratio of areas form a plateau with a concave center. There is no violation  of the triangle inequality. The plots range is  $\beta,\gamma\in \{0,\pi/2\}$.}
\end{figure}

To determine the area of the triangle formed by $A$, $B$ and $C$, we use the definition in Eq.~\ref{eq:area}. We have all the entropies and the conditional entropies we need for this calculation by using the chain rule for multiple random variables, 
\begin{equation}
\label{eq:chain}
H_{ABC} = \underbrace{H_A+H_{B|A}}_{H_{AB}}+H_{C|AB}
\end{equation}
to solve for $H_{C|AB}$.  
The information triangle area ${\mathcal A}_{ABC}^{|GHZ\rangle}$  can be  obtained by Eq,~\ref{eq:area} using our expressions for the joint entropies.  Since the information area  is an involved function of the two detector angles $\beta$ and $\gamma$, we will not display this explicitly. However we evaluate it numerically, and show our results in Fig.~\ref{fig:GHZ}.  It is interesting to us that this entangled state, there is a relatively large region where the  Euclidean area is close to the information area. The information area is well behaved with a local maximum at $\beta=\gamma=\pi/4$.  At this particular maximum, the geometry of the triangle is and isosceles triangle and is embeddable in the Euclidean plane.  Its embedded area ($3$) achieves the upper bound for the area formula in Eq.~\ref{eq:areabounds}  and is different from the Euclidean area ($\sim 1.732$).  At this local minimum, 
\begin{eqnarray}
{\mathcal D}_{AB} & = & {\mathcal D}_{AC}  =  {\mathcal D}_{BC}= 2,\\
{\mathcal A}_{ABC}^{|GHZ\rangle} &=& 3,
\end{eqnarray}
as illustrated in Fig.~\ref{fig:GHZ2}.

The Euclidean area plotted in the middle box of Fig.~\ref{fig:GHZ2} is defined using Heron's formulae, 
\begin{equation} 
\label{eq:ea}
{\mathcal A}_E = \frac{1}{4} \sqrt{  	({\mathcal D}_{AB}+{\mathcal D}_{AC}-{\mathcal D}_{BC})
						({\mathcal D}_{AB}-{\mathcal D}_{AC}+{\mathcal D}_{BC})
						(-{\mathcal D}_{AB}+{\mathcal D}_{AC}+{\mathcal D}_{BC})
						({\mathcal D}_{AB}+{\mathcal D}_{AC}+{\mathcal D}_{BC})
					}.
\end{equation}
The missing sections of the domain is where the triangle inequality is violated. Eq.~\ref{eq:ea} is ideally suited to detect triangle inequality violations as the radical becomes imaginary. 
Perhaps, this is not so surprising since the three-tangle obtains its maximum permitted value of unity for the $|GHZ\rangle$ state \cite{Coffman:2000}. The tangle is the square of the concurrence. We will look at the $|W\rangle$ state in the next section whose three-tangle is $<1$,  but whose pairwise-tangle is maximal and greater  than the pairwise tangle for the $|GHZ\rangle$ state \cite{Briegel:2001}.

%*************************
\subsection{Quantum State Interrogation: the $|W\rangle$ State}
\label{sec:W}

Following the last two subsections, we briefly outline the information geometry of the triangle shown in Fig:~\ref{fig:QN} for the $|W\rangle$ state,
\begin{equation}
\label{eq:W}
|\Psi\rangle \Longrightarrow |W\rangle = \frac{1}{\sqrt{3}} \left( 
	|\updownarrow\leftrightarrow\leftrightarrow\rangle + 
	|\leftrightarrow\updownarrow\leftrightarrow\rangle + 
	|\leftrightarrow\leftrightarrow\updownarrow\rangle \right).
\end{equation}
We will calculate the three edge lengths using the techniques introduced in Sec.~\ref{sec:G} and applied in Sec:~\ref{sec:S}.  We will also calculate the information area Eq.~\ref{eq:area} for this triangle. For the rest of this section we set $\alpha=0$.   

Again we consider the same three observers Alice ($A$), Bob ($B$) and Charlie ($C$). $A$, $B$ and $C$ measure the $|W\rangle$  state with measurement operators given in Eq.~\ref{eq:mo}.  We also set $\alpha=0$ for the remainder of the section.  

The eight joint probabilities from the three measurements on this entangled state $M_C M_B M_A |W\rangle$ are:
\begin{align}
\label{eq:jointW}
\begin{array}{ll}
p(A=1,B=1,C=1) = \frac{1}{3} \sin^2(\beta)\sin^2(\gamma), &  p(A=0,B=1,C=1) =  \frac{1}{3}\sin^2(\beta+\gamma), \\%
p(A=1,B=1,C=0) =  \frac{1}{3}\sin^2(\beta)\cos^2(\gamma), & p(A=0,B=1,C=0) =  \frac{1}{3}\cos^2(\beta+\gamma), \\%
p(A=1,B=0,C=1) =  \frac{1}{3}\cos^2(\beta)\sin^2(\gamma), & p(A=0,B=0,C=1) =  \frac{1}{3}\cos^2(\beta+\gamma), \\%
p(A=1,B=0,C=0) =  \frac{1}{3}\cos^2(\beta)\cos^2(\gamma), & p(A=0,B=0,C=0) =  \frac{1}{3}\sin^2(\beta+\gamma).\\%
\end{array}
\end{align}
Tracing these joint probability over each observer yields the pairwise joint probabilities, 
\begin{align}
\begin{array}{ll}
p(A=1,B=1) &= \frac{1}{3}\sin^2(\beta),\\%
p(A=1,B=0) &= \frac{1}{3}\cos^2(\beta), \\%
p(A=1,C=1) &= \frac{1}{3}\sin^2(\gamma),\\%
p(A=1,C=0) &= \frac{1}{3}\cos^2(\gamma),\\%
p(A=0,B=1) &= \frac{1}{3},\\%
p(A=0,B=0) &= \frac{1}{3}, \\%
p(A=0,C=1) &= \frac{1}{3},\\ %
p(A=0,C=0) &= \frac{1}{3},\\ %
p(B=1,C=1) &= \frac{1}{3}\left(\sin^2(\beta)\sin^2(\gamma) + \sin^2(\beta+\gamma) \right), \\ %
p(B=1,C=0) &= \frac{1}{3}\left( \sin^2(\beta)\cos^2(\gamma) + \cos^2(\beta+\gamma) \right), \\ %
p(B=0,C=1) &=  \frac{1}{3}\left( \cos^2(\beta)\sin^2(\gamma) + \cos^2(\beta+\gamma) \right), \\ %
p(B=0,C=0) &=  \frac{1}{3}\left( \cos^2(\beta)\cos^2(\gamma) +\sin^2(\beta+\gamma) \right). %
\end{array}
\end{align}
Finally, tracing the joint probability over all pairs of observers gives us the six probabilities for the measurement outcomes of $A$, $B$ and $C$, 
\begin{equation} \label{eq:w1}
\begin{array}{lr}
p(A=0) = 2/3, & p(A=1) = 1/3 \\ %
p(B=0) = 2/3, & p(B=1) = 1/3 \\ %
p(C=0) = 2/3, & p(C=1) = 1/3 \\ %
\end{array}.
\end{equation} 

The pairwise conditional probabilities can be recovered from these pairwise joint probabilities since using Eq~\ref{eq:joint}.

We are now in a position to use Eqs.~\ref{eq:entropy}-\ref{eq:centropy} to calculate the entropy, the conditional entropy as well as the information distance in Eq.~\ref{eq:length}.  The entropy of our observers are all equal, 
\begin{equation}
H_A = H_B = H_C = \log{(3)} - \frac{2}{3}.
\end{equation}
The joint entropy between pairs of our observers are,
\begin{equation}
\label{eq:we1}
\begin{array}{rl}
H_{AB} = &  \log(3) - \frac{1}{3} \sin^2(\beta) \log(\sin^2(\beta)) -\frac{1}{3} \cos^2(\beta) \log(\cos^2(\beta)),\\
H_{AC} = &  \log(3) - \frac{1}{3} \sin^2(\gamma) \log(\sin^2(\gamma)) -\frac{1}{3} \cos^2(\gamma) \log(\cos^2(\gamma)),\\
H_{BC} = &  \log(3) -  \frac{1}{3}\left(\sin^2(\beta)\sin^2(\gamma) + \sin^2(\beta+\gamma) \right) \log\left(\sin^2(\beta)\sin^2(\gamma) + \sin^2(\beta+\gamma) \right)\\
                 & -\frac{1}{3}\left( \sin^2(\beta)\cos^2(\gamma) + \cos^2(\beta+\gamma) \right) \log\left( \sin^2(\beta)\cos^2(\gamma) + \cos^2(\beta+\gamma) \right)\\
                 & -\frac{1}{3}\left( \cos^2(\beta)\sin^2(\gamma) + \cos^2(\beta+\gamma) \right) \log\left( \cos^2(\beta)\sin^2(\gamma) + \cos^2(\beta+\gamma) \right)\\
                 & -\frac{1}{3}\left( \cos^2(\beta)\cos^2(\gamma) +\sin^2(\beta+\gamma) \right) \log\left( \cos^2(\beta)\cos^2(\gamma) +\sin^2(\beta+\gamma) \right).
\end{array}
\end{equation}
Finally, we use Eq.~\ref{eq:joint} to find the joint entropy $H_{ABC}$ of $A$, $B$ and $C$
\begin{equation}
\label{eq:jeW}
\begin{array}{rl}
H_{ABC} =& \log(3)-\frac{1}{3} \sin^2(\beta)\sin^2(\gamma)\log(\sin^2(\beta)\sin^2(\gamma)) - \frac{1}{3}\sin^2(\beta)\cos^2(\gamma)\log(\sin^2(\beta)\cos^2(\gamma)) \\
                  & -\frac{1}{3}\cos^2(\beta)\sin^2(\gamma)\log(\cos^2(\beta)\sin^2) -\frac{1}{3}\cos^2(\beta)\cos^2(\gamma)\log(\cos^2(\beta)\cos^2(\gamma)) \\
                  & -\frac{1}{3}\sin^2(\beta+\gamma)\log(\sin^2(\beta+\gamma))-\frac{1}{3}\cos^2(\beta+\gamma)\log(\cos^2(\beta+\gamma)) \\
                  & -\frac{1}{3}\cos^2(\beta+\gamma)\log(\cos^2(\beta+\gamma)) -\frac{1}{3}\sin^2(\beta+\gamma)\log(\sin^2(\beta+\gamma)) .
                \end{array}
\end{equation}

The three lengths of the edges of the triangle for this $|W\rangle$ state are derived from Eq.~\ref{eq:length}, and are
\begin{equation}
\label{eq:3lw}
\begin{array}{rl}
{\mathcal D}_{AB}  = &  \frac{1}{3} - \frac{2}{3}\sin^2{(\beta)} \log\left(\sin^2{(\beta)}\right) -\frac{2}{3} \cos^2{(\beta)} \log\left(\cos^2{(\beta)}\right),\\
{\mathcal D}_{AC}  = & \frac{1}{3} - \frac{2}{3}\sin^2{(\gamma)} \log\left(\sin^2{(\gamma)}\right) -\frac{2}{3} \cos^2{(\gamma)} \log\left(\cos^2{(\gamma)}\right),\\
{\mathcal D}_{BC}  = & \frac{1}{3}-\frac{2}{3} \sin^2(\beta)\sin^2(\gamma)\log(\sin^2(\beta)\sin^2) - \frac{2}{3}\sin^2(\beta)\cos^2(\gamma)\log(\sin^2(\beta)\cos^2(\gamma)) \\
                  & -\frac{2}{3}\cos^2(\beta)\sin^2(\gamma)\log(\cos^2(\beta)\sin^2) -\frac{2}{3}\cos^2(\beta)\cos^2(\gamma)\log(\cos^2(\beta)\cos^2(\gamma)) \\
                  & -\frac{2}{3}\sin^2(\beta+\gamma)\log(\sin^2(\beta+\gamma))-\frac{2}{3}\cos^2(\beta+\gamma)\log(\cos^2(\beta+\gamma)) \\
                  & -\frac{2}{3}\cos^2(\beta+\gamma)\log(\cos^2(\beta+\gamma)) -\frac{2}{3}\sin^2(\beta+\gamma)\log(\sin^2(\beta+\gamma)).
\end{array}
\end{equation}

\begin{figure} [ht]
\centering
   \includegraphics[height=2 in]{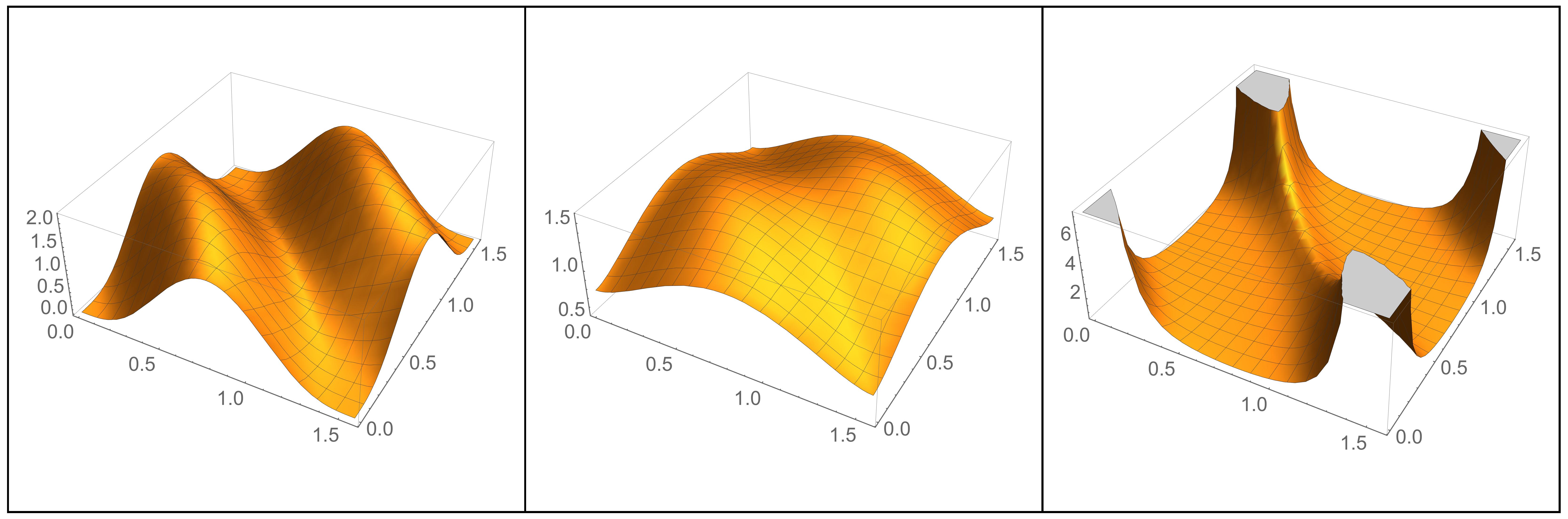}
   \caption
   { \label{fig:w} 
The left plot is the information area ${\mathcal A}_{ABC}^{|W\rangle}(\beta,\gamma)$, the center plot is the Euclidean area (${\mathcal A}_E$) based on the three information distances in Eq.~\ref{eq:3lw}, and the right plot is the ratio, where we see a diagonal ridge line. The plots range is  $\beta,\gamma\in \{0,\pi/2\}$.}
   \end{figure}

To determine the area of the triangle formed by $A$, $B$ and $C$ we use the definition in Eq.~\ref{eq:area}. We have all the entropies, conditional entropies, and chain rule for multiple random variables in Eq.~\ref{eq:chain} to solve for $H_{C|AB}, etc.$.  
The information triangle area ${\mathcal A}_{ABC}^{|W\rangle}$  can be  obtained by Eq,~\ref{eq:area} using our expressions for the joint entropies.  As it is an involved function of the two detector angles $\beta$ and $\gamma$, we will not display this explicitly. However we evaluate it numerically and illustrate our results in Fig.~\ref{fig:w}.
It is interesting that this entangled state is significantly different than the $|GHZ\rangle$ state.  It has a saddle point rather than a local minimum. The information area is well behaved with a local saddle point at $\alpha=\beta=\pi/4$.  The Euclidean area based on Eq.~\ref{eq:ea} is well defined over the entire range indicating that the information triangles never violate the triangle inequality. At the saddle point in the geometry,  the triangle is embeddable in the Euclidean plane and its embedded area  ($1.362$) is substantially larger than the information area indicating curvature. At this point, 
\begin{align}
{\mathcal D}_{AB} & = {\mathcal D}_{AC} = 2,\\
{\mathcal D}_{BC} &= 1.463, and\\
{\mathcal A}_{ABC}^{|W\rangle} &= 0.512.
\end{align}
as illustrated in Fig.~\ref{fig:w}.
It would be interesting to see if this is related to the large two-way tangle for this state. We can conclude two things thus far, (1) triangle inequality violations do not appear here, and (2) the information area of the $W\rangle$ state is qualitatively different from the $|GHZ\rangle$ state. Obviously, more needs to be done to measure entanglement. 

%*************************
\subsection{Quantum State Interrogation: a Separable State $|P\rangle$}
\label{sec:P}

Following the last two subsections, we briefly outline  the information geometry of the triangle shown in Fig:~\ref{fig:QN} for the separable state,
\begin{equation}
\label{eq:P}
|\Psi\rangle \Longrightarrow |P\rangle = |\updownarrow\updownarrow\updownarrow\rangle. 
\end{equation}
We calculate the three edge lengths using the techniques introduced in Sec.~\ref{sec:G} and applied in Sec:~\ref{sec:S}.  We will also calculate the information area for this triangle that is given by Eq.~\ref{eq:area}.   

Again we consider three observers Alice ($A$), Bob ($B$) and Charlie ($C$). $A$, $B$ and $C$ measure the separable state using their choice of detectors defined in Eq.~\ref{eq:mo}.  For the rest of this section we also set $\alpha=0$.  

The only four non-vanishing joint probabilities from the three measurements on this separable state $M_C M_B M_A |P\rangle$ are,
\begin{align}
\label{eq:jointP}
p(A=1,B=1,C=1) &= \cos^2(\beta)\cos^2(\gamma), \\
p(A=1,B=1,C=0) &=  \cos^2(\beta)\sin^2(\gamma), \\
p(A=1,B=0,C=1) &=  \sin^2(\beta)\cos^2(\gamma), \\
p(A=1,B=0,C=0) &=  \sin^2(\beta)\sin^2(\gamma). 
\end{align}
Tracing the joint probability successively over the three observers gives us the eight non-vanishing pairwise joint probabilities,
\begin{align}
p(A=1,B=1) &= \cos^2(\beta),\\
p(A=1,B=0) &= \sin^2(\beta), \\
p(A=1,C=1) &= \cos^2(\gamma),\\
p(A=1,C=0) &= \sin^2(\gamma),\\
p(B=1,C=1) &= \cos^2(\beta)\cos^2(\gamma), \\
p(B=1,C=0) &= \cos^2(\beta)\sin^2(\gamma), \\
p(B=0,C=1) &=  \sin^2(\beta)\cos^2(\gamma), \\
p(B=0,C=0) &=  \sin^2(\beta)\sin^2(\gamma). 
\end{align}
Finally, tracing the joint probability over all pairs of observers gives us the six probabilities for the measurement outcomes of $A$, $B$ and $C$,
\begin{equation} 
\label{eq:p1P}
\begin{array}{lr}
p(A=0) = 0, & p(A=1) = 1, \\
p(B=0) = \sin^2{(\beta)},  & p(B=1) = \cos^2{(\beta)}, \\
p(C=0) = \sin^2{(\gamma)},  & p(C=1) = \cos^2{(\gamma)}.  \\
\end{array}
\end{equation}
The conditional probabilities can be recovered from these probabilities using Eq.\ref{eq:joint} and we find the following three sets of conditional probabilities ($A$-$B$, $A$-$C$ and $B$-$C$):
\begin{equation} \label{eq:pABP}
\begin{array}{|c|c|}
\hline
p(A=0|B=0) = 0 & p(A=1|B=0) = 1 \\
\hline
p(A=0|B=1) = 0 & p(A=1|B=1) = 1 \\
\hline
p(B=0|A=0) = NA & p(B=1|A=0) = NA \\
\hline
p(B=0|A=1) = \sin^2{(\beta)} & p(B=1|A=1) = \cos^2{(\beta)} \\
\hline
\end{array};
\end{equation} 
\begin{equation} \label{eq:pACP}
\begin{array}{|c|c|}
\hline
p(A=0|C=0) = 0 & p(A=1|C=0) = 1\\
\hline
p(A=0|C=1) = 0 & p(A=1|C=1) = 1 \\
\hline
p(C=0|A=0) = NA & p(C=1|A=0) = NA \\
\hline
p(C=0|A=1) = \sin^2{(\gamma)} & p(C=1|A=1) = \cos^2{(\gamma)} \\
\hline
\end{array};
\end{equation} 
\begin{equation} \label{eq:pBCP}
\begin{array}{|c|c|}
\hline
p(B=0|C=0) = \sin^2{(\beta)} & p(B=1|C=0) = \cos^2{(\beta)} \\
\hline
p(B=0|C=1) = \sin^2{(\beta)} & p(B=1|C=0) = \cos^2{(\beta)} \\
\hline
p(C=0|B=0) = \sin^2{(\gamma)} & p(C=1|B=0) = \cos^2{(\gamma)} \\
\hline
p(C=0|B=1) = \sin^2{(\gamma)} & p(C=1|B=1) = \cos^2{(\gamma)} \\

\hline
\end{array}.
\end{equation} 

We are now in a position to use Eqs.~\ref{eq:entropy}-\ref{eq:centropy} to calculate the entropy, the conditional entropy as well as the information distance in Eq.~\ref{eq:length}.  The entropy of our observers are
\begin{align}
H_A &= 0,\\
H_B &= -\sin^2(\beta) \log(\sin^2(\beta)) -\cos^2(\beta) \log(\cos^2(\beta)),\\
H_C &= -\sin^2(\gamma) \log(\sin^2(\gamma)) -\cos^2(\gamma) \log(\cos^2(\gamma)),\\
\end{align}
and the joint entropy between pairs of our observers are,
\begin{equation}
\label{eq:pe1}
\begin{array}{rl}
H_{AB} = &  -\sin^2(\beta) \log(\sin^2(\beta)) -\cos^2(\beta) \log(\cos^2(\beta)),\\
H_{AC} = & -\sin^2(\gamma) \log(\sin^2(\gamma)) -\cos^2(\gamma) \log(\cos^2(\gamma)),\\
H_{BC} = & -\sin^2{(\gamma)}\sin^2{(\beta)} \log(\sin^2{(\gamma)}\sin^2{(\beta)}) - \cos^2{(\gamma)}\sin^2{(\beta)} \log(\cos^2{(\gamma)}\sin^2{(\beta)}) \\
& -\sin^2{(\gamma)}\cos^2{(\beta)} \log(\sin^2{(\gamma)}\cos^2{(\beta)}) - \cos^2{(\gamma)}\cos^2{(\beta)} \log(\cos^2{(\gamma)}\cos^2{(\beta)}).
\end{array}
\end{equation}
Finally, we use Eq.~\ref{eq:joint} to find the joint entropy $H_{ABC}$ of $A$, $B$ and $C$, it simplifies to
\begin{eqnarray}
\label{eq:jep}
H_{ABC} = H_{BC},
\end{eqnarray}
where $H_{BC}$ is given explicitly in Eq.~\ref{eq:pe1}.
 We find the three lengths of the edges of the triangle for this separable state from Eq.~\ref{eq:length},
\begin{equation}
\label{eq:3lp}
\begin{array}{rl}
{\mathcal D}_{AB} = &  -\sin^2(\beta) \log(\sin^2(\beta)) -\cos^2(\beta) \log(\cos^2(\beta)),\\
{\mathcal D}_{AC} = & -\sin^2(\gamma) \log(\sin^2(\gamma)) -\cos^2(\gamma) \log(\cos^2(\gamma)),\\
{\mathcal D}_{BC} = & -2\sin^2{(\gamma)}\sin^2{(\beta)} \log(\sin^2{(\gamma)}\sin^2{(\beta)}) - 2\cos^2{(\gamma)}\sin^2{(\beta)} \log(\cos^2{(\gamma)}\sin^2{(\beta)}) \\
& -2 \sin^2{(\gamma)}\cos^2{(\beta)} \log(\sin^2{(\gamma)}\cos^2{(\beta)}) - 2 \cos^2{(\gamma)}\cos^2{(\beta)} \log(\cos^2{(\gamma)}\cos^2{(\beta)})  \\
& -\sin^2(\beta) \log(\sin^2(\beta)) -\cos^2(\beta) \log(\cos^2(\beta)) -\sin^2(\gamma) \log(\sin^2(\gamma)) -\cos^2(\gamma) \log(\cos^2(\gamma)).
\end{array}
\end{equation}
The area of the triangle for this separable state using  Eq.~\ref{eq:area} and  Eq.~\ref{eq:jep} is, 
\begin{equation}
\label{eq:areaP}
{\mathcal A}_{ABC}^{|P\rangle}\left(\beta,\gamma\right) =  H_{BC}^2 -H_{AB}H_{BC}-H_{AC}H_{BC}+H_{AB}H_{AC}.
\end{equation}

\begin{figure}
   \centering
  % \begin{tabular}{c} %% tabular useful for creating an array of images 
   \includegraphics[height=2 in]{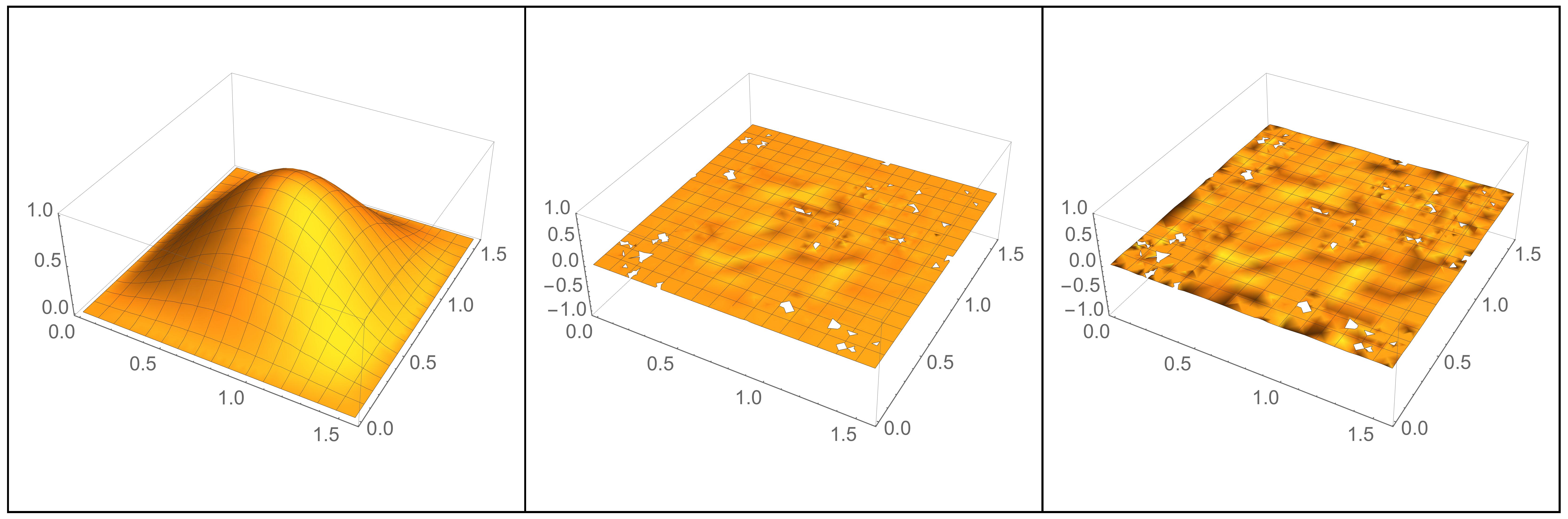}
  % \end{tabular}
  % \end{center}
   \caption{\label{fig:p}
 	The left plot is the information area ( ${\mathcal A}_{ABC}^{|P\rangle}(\beta,\gamma)$), the center plot is the Euclidean area (${\mathcal A}_E$) based on the three information distances in Eq.~\ref{eq:3lp}, and the right 	plot is the ratio, where we see $ {\mathcal A}_E \sim e {\mathcal A}_{ABC}$. The plots range is $\beta,\gamma\in \{0,\pi/2\}$.}
   \end{figure}

It is interesting that for this separable state, the Euclidean area is essentially zero for all values of $\beta$ and $\gamma$. The information area is well behaved with a maximum at $\beta=\gamma=\pi/4$.  In this particular case the geometry of the triangle collapses in the Euclidean plane to a line. At this global maximum in information area we find, 
\begin{align}
{\mathcal D}_{AB} & = {\mathcal D}_{AC} = 1,\\
{\mathcal D}_{BC} &= {\mathcal D}_{AB}+{\mathcal D}_{AC}=2, and\\
{\mathcal A}_{ABC}^{|P\rangle} &= 1.
\end{align}
 This is illustrated in Fig.~\ref{fig:p}.

%*************************
\section{From Bipartite Quadrilateral to Tripartite Octagon: Future Prospects}
\label{sec:O}

What we outlined in this manuscript is the beginning of our exploration of our approach to information geometry.  We do not have a definitive entanglement measure or actual scalability results. However, the analysis of the triangles for the $|GHZ\rangle$, $|W\rangle$ and separable state $|P\rangle$ yielded qualitatively unique features. The relative differences between the perimeters of each of the triangles and the corresponding information area suggests that a curvature measure might be useful for differentiating quantum states, and in particular the separable state $|\updownarrow\updownarrow\updownarrow\rangle$ showed an interesting feature of nearly zero area from Heron's formulae. We also observed that the triangle inequality was not violated for these entangled states in the measurement space we considered. In particular, there were no violations for the $|GHZ\rangle$ and, we observed that there were never triangle inequality violations for the $|W\rangle$ state. It is clear that an exhaustive exploration is needed, and this seems feasible. We have the six Bloch sphere detector angles to explore, as well as the need to explore a sampling of the space of symmetric tripartite states. It is equally clear that new measures and ideas are needed. We discuss a promising avenue in the remainder of this section. 

\begin{figure} [ht]
\centering
   \includegraphics[height=3.0in]{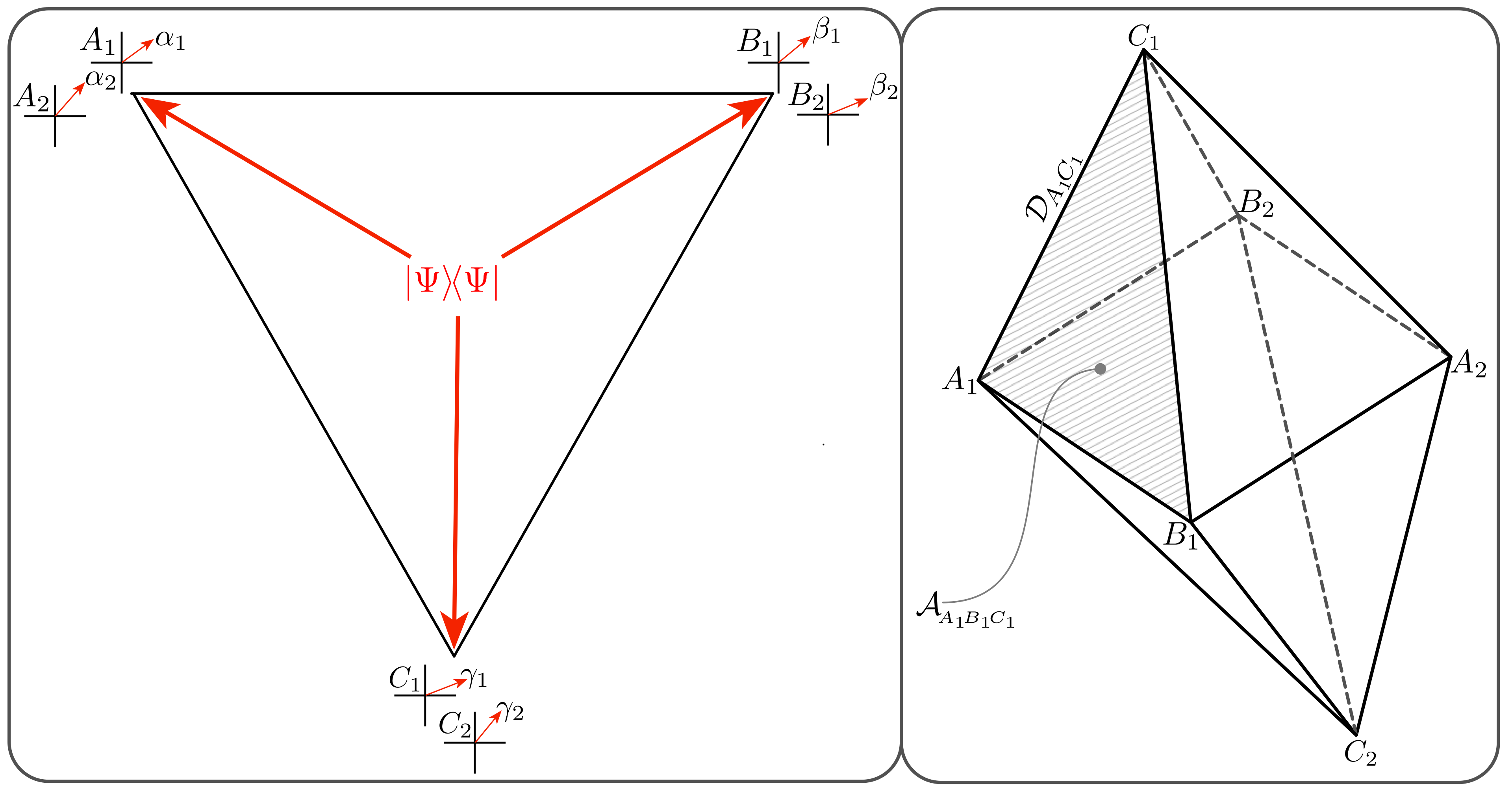}
   \caption
   { \label{fig:OG} 
This figure illustrates a generalization of Schumacher's quantum information geometry of a bipartite state to our information geometry of a tripartite state.   Each of the three observers have two detectors (left box). The quadrilateral of Schumacher (Fig~\ref{fig:S}) generalizes naturally to an octagon (right box).  We can use our formalism to study its information geometry. We can use this arrangement to explore some curvature and embeddability properties of the octahedron as indicators of the properties of the quantum state, $|\Psi\rangle$.}
   \end{figure}

We have generalized the Schumacher approach from bipartite states to tripartite states.  If each of our observers, $A$, $B$ and $C$ can choose randomly between two separate detectors then the triangle becomes an octahedron as illustrated in Fig.~\ref{fig:OG}. For the bipartite system each of the two observers had two detectors so the line connecting the observer became the quadrilateral illustrated in the lower half of Fig.~\ref{fig:S}. In our generalization to tripartite states, each of our three observers has two detectors, and the triangle connecting them becomes an octahedron as shown in the right box of Fig.~\ref{fig:OG}.
In this configuration we can calculate the 12  lengths of the edges of the octahedron, we can also calculate the 8 triangle areas. We can then ask questions as to the embeddability of the octahedron in Euclidean and Minkowski space, its curvature, and other measures. This work is in progress and will be reported at a later date. 

%\appendix    %>>>> this command starts appendixes

\section*{ACKNOWLEDGMENTS}       
 
This research benefited form discussions with P. M. Alsing and his group, and from discussions with M. Corne and  S. Mostafanazhad Aslmarand.  We thank AFRL/RITA and the Griffiss Institute  for providing a stimulating research environment and support under the Summer Faculty Fellowship Program. This research was supported under AFOSF/AOARD grant \#FA2386-17-1-4070.      

\bibliography{report} 
\bibliographystyle{amsplain} 

\end{document}